\documentclass[%
 aip,
amsmath,amssymb,
reprint,%
]{revtex4-1}

\usepackage{graphicx}
\usepackage{dcolumn}
\usepackage{bm}

\usepackage[utf8]{inputenc}
\usepackage[T1]{fontenc}
\usepackage{mathptmx}
\usepackage{etoolbox}
\usepackage{afterpage}
\usepackage{xcolor}  

\makeatletter
\def\@email#1#2{%
 \endgroup
 \patchcmd{\titleblock@produce}
  {\frontmatter@RRAPformat}
  {\frontmatter@RRAPformat{\produce@RRAP{*#1\href{mailto:#2}{#2}}}\frontmatter@RRAPformat}
  {}{}
}%
\makeatother
\begin{document}

\preprint{AIP/123-QED}

\title[The interplay of plasticity and elasticity in elastoviscoplastic flows in wavy channels]{The interplay of plasticity and elasticity in elastoviscoplastic flows in wavy channels}
\author{Mohamed S. Abdelgawad}
 \affiliation{ 
Complex Fluids and Flows Unit, Okinawa Institute of Science and Technology Graduate University, 1919-1 Tancha, Onna-son, Okinawa 904-0495, Japan
}%
\author{Simon J. Haward}%
\affiliation{ 
Micro/Bio/Nanofluidics Unit, Okinawa Institute of Science and Technology Graduate University, 1919-1 Tancha, Onna-son, Okinawa 904-0495, Japan
}%

\author{Amy Q. Shen}
\affiliation{ 
Micro/Bio/Nanofluidics Unit, Okinawa Institute of Science and Technology Graduate University, 1919-1 Tancha, Onna-son, Okinawa 904-0495, Japan
}%

\author{Marco E. Rosti}
 \homepage{Author to whom correspondence should be addressed: marco.rosti@oist.jp}
\affiliation{ 
Complex Fluids and Flows Unit, Okinawa Institute of Science and Technology Graduate University, 1919-1 Tancha, Onna-son, Okinawa 904-0495, Japan
}%

\date{\today}

\begin{abstract}
Elastoviscoplastic (EVP) fluids, which exhibit both solid-like and liquid-like behavior depending on the applied stress, are critical in industrial processes involving complex geometries such as porous media and wavy channels. In this study, we investigate how flow characteristics and channel design affect EVP fluid flow through a wavy channel, using numerical simulations supported by microfluidic experiments. Our results reveal that elasticity significantly influences flow dynamics, reducing pressure drops and expanding unyielded regions. Notably, we find that even minimal elasticity can shift the flow from steady to time-dependent regimes, a transition less pronounced in viscoelastic fluids. Additionally, we show that the development of stagnation regions can be prevented when using a modified EVP fluid with enhanced elasticity, thus providing a full global yielding of the material. This study elucidates the role of elasticity in modifying flow patterns and stress distribution within EVP fluids, offering insights into the optimization of industrial applications, such as the displacement of yield stress fluids in enhanced oil recovery, gas extraction, cementing, and other processes where flow efficiency is critical.
\end{abstract}

\maketitle

\section{\label{sec:level1}Introduction}
Elastoviscoplastic (EVP) fluids represent a unique class of non-Newtonian fluids, exhibiting complex rheological behavior characterized by a combination of elastic, viscous, and plastic properties. These fluids possess a yield stress, a critical property that determines their transition from a solid-like to a liquid-like state: in particular, they maintain a solid-like behavior below this yield stress, but once this threshold is surpassed, they flow like liquids~\cite{Balmforth2014, Coussot2014}. Importantly, their elastic characteristics persist in both states, allowing them to store energy and resist deformations~\cite{Dinkgreve2017, Malkin2017, Gouamba2019}.

EVP fluids include products like toothpaste, ketchup, and cosmetics, and are crucial in industries ranging from construction to oil extraction~\cite{Balmforth2014, Coussot2014}. These fluids often navigate complex geometries, such as wavy channels and porous media, in various geophysical and industrial applications. One notable example is in oil recovery, where heavy crude oils, which also exhibit yield stress and viscoelastic properties, must move through intricate porous structures and curved pathways~\cite{barenblatt1990}. In oil and gas well construction, encountering varied geological strata can lead to uneven drilling conditions and washout formations. Drilling muds, characterized by their yield stress, are essential here; they are pumped down the drill pipe to cool the drill bit and remove cuttings, circulated back to the surface. The cement slurry, pumped under pressure inside the casing, must navigate these washouts, making its way around the borehole and back up, displacing the drilling mud. This uneven spacing between the casing and borehole wall can cause drilling mud to stagnate, complicating the cementing process~\cite{Nelson2006}. Further examples of EVP fluid flows in complex geometries include the injection of cement for effective CO$_2$ storage sealing, sealants for moisture control in building brickwork, and fracking fluids to expand wellbore cracks and stimulate oil and gas production~\cite{Barati2014,Zhu2021}. 

Extensive studies on inelastic Bingham fluids -- a kind of viscoplastic (VP) material -- have investigated the effect of channel geometry and flow conditions in wavy-walled channels~\cite{Roustaei2013, Roustaei2015a, Roustaei2015b, Roustaei2016}. For instance, in channels with minimal amplitude and elongated wavelength perturbations, the central plug region maintains its rigidity. However, increasing the amplitude causes the plug to break apart, resulting in a combination of rigid and low shear extensional pseudo-plug regions at the channel center~\cite{Roustaei2013}. Once the amplitude of channel wall perturbations exceeds a critical threshold, the fluid may stagnate in the widest section of the channel, creating stationary zones known as fouling layers~\cite{Roustaei2013}. The emergence of these stationary regions stems largely from the channel geometry and the extensional stresses induced by wall structures~\cite{Roustaei2015a}. With significant wall perturbations and high fluid plasticity, the yielded region becomes geometry-independent. Notably, this self-selection of the yielded region persists even when considering inertial effects ~\cite{Roustaei2015b}. Moreover, the pressure drop across the channel is influenced by wall perturbations and fluid plasticity, approaching a constant value when the yielded region assumes a fixed shape~\cite{Roustaei2016}. These studies reveal that the classical Darcy's law fails to adequately describe the flow rate-pressure drop relationship in uneven-walled channels~\cite{Roustaei2016}. However, it is important to note that, these findings pertain only to VP fluids, where the effects of elasticity were not considered.

The significance of incorporating elasticity in simulations of yield stress fluids is driven by the experimental evidence of elastic effects in such flows. These effects manifest as fore-aft asymmetry and negative wakes behind settling particles~\cite{Holenberg2012,Fraggedakis2016}, as well as the cuspid shape of bubbles rising in yield stress fluids~\cite{daneshi_frigaard_2023,Moschopoulos2021}. Elasticity also contributes to the asymmetry of the yield surface, particularly through abrupt changes in cross-sectional areas, as observed in experimental studies with Carbopol solutions ~\cite{Varges2020, Villalba2023}. Consequently, it is crucial to account for both elastic and plastic effects in yield stress fluids to accurately predict flow dynamics~\cite{Saramito2007,Saramito2009}. Numerical models, particularly elastoviscoplastic ones, are designed to incorporate these effects, enhancing predictions of flow behavior and the shape of unyielded regions~\cite{Fraggedakis2016b}. Furthermore, studies on EVP flows through porous media indicate that elasticity not only increases the pressure drops, but also enlarges unyielded regions and introduces time-dependent flow behaviors, which are not typically observed in purely viscoplastic flows~\cite{DeVita2018,Chaparian2020,Parvar2024}.

In this study, we examine the flow of EVP fluids through wavy channels, specifically investigating how elasticity and plasticity, along with the channel geometry, influence bulk flow behaviors, such as pressure drop, variations in yielded and unyielded regions, and flow regimes. While previous research has extensively explored VP fluid behaviors, the combined effects of elasticity and plasticity, particularly in complex geometries like wavy channels, have been largely overlooked.  By addressing this gap, our work aims to provide a comprehensive understanding of the interplay between elasticity and yield stress in complex flow environments. 

\section{Governing equations and flow geometry}

The dynamics of EVP fluid flow through a wavy channel is governed by the Navier–Stokes equations, coupled with the constitutive equation for the non-Newtonian extra stress tensor. We consider incompressible flows, thus enforcing that the divergence of the fluid velocity vector $\boldsymbol{u}$ is zero,
\begin{equation}
\label{eq:mass}
	\nabla \cdot \boldsymbol{u}=0.
\end{equation}
The momentum conservation equation is expressed as 
\begin{equation}
\label{eq:mom}
 \rho \left( \frac{\partial \boldsymbol{u}}{\partial t} + \boldsymbol{u} \cdot \nabla \boldsymbol{u} \right) = -\nabla p + \nabla \cdot \boldsymbol{\sigma} + \boldsymbol{f},
\end{equation}
where $\rho$ is the fluid density and $p$ the pressure field. The total stress tensor $\boldsymbol{\sigma}$ incorporates both the Newtonian viscous stress and the non-Newtonian extra stress tensor $\boldsymbol{\tau}$: $\boldsymbol{\sigma} = 2 \eta_{s} \boldsymbol{\dot{\gamma}} + \boldsymbol{\tau}$, where $\eta_s$ denotes the solvent viscosity, and $\boldsymbol{\dot{\gamma}}$ the rate of deformation tensor defined as: $\boldsymbol{\dot{\gamma}} = \left( \nabla\boldsymbol{u} + \left(\nabla\boldsymbol{u}\right)^{T} \right)/2$. The external body force $\boldsymbol{f}$, applied through an immersed boundary method~\cite{HORI2022}, enforces the boundary conditions directly within the flow field, ensuring an accurate representation of flow interactions with solid boundaries.

\begin{figure}[!h]
\includegraphics[width = 0.4\textwidth]{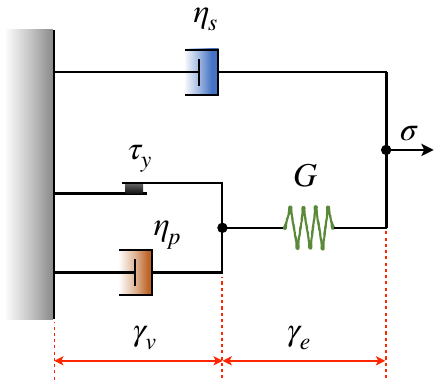}
\caption{\label{fig:SRM} Schematic of the mechanical representation of the Saramito model.}
\end{figure}

For the non-Newtonian stress tensor $\boldsymbol{\tau}$, we adopt the constitutive model proposed by \citet{Saramito2007} (SRM), which introduces elasticity into the classical Bingham model. A one-dimensional schematic of the mechanical representation of the model is illustrated in Fig.~\ref{fig:SRM}.
In this model, $\tau_y$ represents the yield stress through a friction element, $G$ is the elastic modulus modeled as a spring, and $\sigma$ denotes the total stress represented by a force that drives the system. The solvent viscosity $\eta_{s}$ and the polymeric viscosity $\eta_{p}$ are depicted as dash-pots. 
Below the yield stress threshold, the friction element is rigid, the material then behaves as a Kelvin-Voigt viscoelastic solid, characterized by the spring $G$ and the viscous element $\eta_s$. Exceeding this threshold, the friction element yields, allowing deformation of all elements including $\eta_{p}$, with the material then behaving like an Oldroyd-B viscoelastic fluid.

The constitutive equation of the SRM model assumes that the total deformation rate $\boldsymbol{\dot{\gamma}}$ is the sum of the elastic deformation rate $\boldsymbol{\dot{\gamma}_{e}}$ and the viscoplastic deformation rate $\boldsymbol{\dot{\gamma}_{v}}$:   
\begin{equation}\label{eq:SRM}
\underbrace{\frac{1}{G}\overset{\nabla}{\boldsymbol{\tau}}}_{\boldsymbol{\dot{\gamma}_{e}}}+\underbrace{\max\left(0,\frac{\tau_{d}-\tau_{y}}{\eta_{p} \tau_{d}}\right)\boldsymbol{\tau}}_{\boldsymbol{\dot{\gamma}_{v}}}=2 \boldsymbol{\dot{\gamma}},
\end{equation}
where $\overset{\nabla}{\boldsymbol{\tau}}$ is the upper convected derivative of the EVP extra stress tensor, calculated as: $\overset{\nabla}{\boldsymbol{\tau}}\equiv\partial\boldsymbol{\tau}/\partial t+\boldsymbol{u}\cdot\nabla\boldsymbol{\tau}-(\nabla\boldsymbol{u})^{T}\cdot\boldsymbol{\tau}-\boldsymbol{\tau}\cdot\nabla\boldsymbol{u}$,
and $\tau_{d}$ is the magnitude of the deviatoric part of the EVP stress tensor ${\tau}_{d}$, defined as:
$\tau_{d} = \sqrt{(\boldsymbol{\tau}_d:\boldsymbol{\tau}_d)/2}$, where $\boldsymbol{\tau}_{d} = \boldsymbol{\tau}- \mathrm{tr}(\boldsymbol{\tau})\boldsymbol{I}/3$, with $\boldsymbol{I}$ being the identity tensor. 
The key difference between the SRM and Bingham models lies in the potential for deformation below the yield stress in the SRM model, attributed to the persistent contribution from the elastic component. Conversely, traditional yield stress fluids modeled by the Bingham equation do not deform until the yield stress threshold is surpassed.
The SRM model has been extended by incorporating elasticity into the classical Herschel-Bulkley model to introduce shear-thinning behavior, resulting in the so-called SRM/HB model~\cite{Saramito2009}:
\begin{equation}\label{eq:SRM/HB}
\frac{1}{G} \overset{\nabla}{\boldsymbol{\tau}}+\max\left(0,\frac{\tau_{d}-\tau_{y}}{k \tau_{d}^{n}}\right)^{\frac{1}{n}}\boldsymbol{\tau}=2\boldsymbol{\dot{\gamma}},
\end{equation}
where $n$ and $k$ are the power and consistency indices, respectively. Thus, the polymeric viscosity is $\eta_p = k \dot{\gamma}^{n}$, where $\dot{\gamma}$ s the magnitude of the rate of deformation tensor.
The SRM/HB constitutive equation (Eq. \ref{eq:SRM/HB}) can describe the shear thickening ($n>1$) or the shear thinning ($n<1$) behaviors observed in many materials post-yielding. Moreover, incorporating two additional parameters ($n$ and $k$) provides more flexibility in fitting experimental data.

A number of dimensionless parameters define the flow characteristics. Considering $D$ and $U$ as the characteristic length and velocity of the problem, we define a characteristic time scale $T$ as $T=D/U$. Similarly, the characteristic stress $S$ is defined as $S= \eta_0 U/D$, with $\eta_0 = \eta_s + \eta_p$ being the total viscosity. Accordingly, the governing equations (Eq.\ref{eq:mass}-\ref{eq:SRM}) can be written in the following dimensionless form,
\begin{equation}\label{eq:mass_dimless}
	\nabla^{*}\cdot \boldsymbol{u}^{*}=0,
\end{equation}
\begin{equation}\label{eq:mom_dimless}
	Re \left(\frac{\partial \boldsymbol{u}^{*}}{\partial t^{*}}+\boldsymbol{u}^{*}\nabla ^{*}\boldsymbol{u}^{*}\right)=-\nabla^{*} p^{*}+\nabla^{*}\cdot\boldsymbol{\sigma}^{*}+\boldsymbol{f}^{*},
\end{equation}
\begin{equation}\label{eq:evp_dimless}
	Wi~\overset{\nabla}{\boldsymbol{\tau}^{*}}+\max\left(0,\frac{\tau_{d}^{*}-Bn}{\tau_{d}^{*}}\right)\boldsymbol{\tau}^{*}=\alpha \boldsymbol{\gamma} ^{*},
\end{equation}
which include the following dimensionless numbers: the Reynolds number ($Re$), which compares inertial to viscous forces and is calculated as $Re = \rho U D/\eta_0$, the Bingham number ($Bn$), indicating the ratio of yield to viscous stresses and is computed as $Bn = \tau_{y}D/\eta_0 U$, the Weissenberg number ($Wi$), comparing elastic to viscous forces and is defined as $Wi = \eta_p U/ G D$, and the viscosity ratio $\alpha = \eta_p/\eta_0$, related to the concentration of polymers. 

\subsection{Numerical methods}

To solve the governing equations numerically, we use the in-house developed flow solver \textit{Fujin} (\texttt{https://groups.oist.jp/cffu/code}), which operates on a staggered uniform Cartesian grid. This configuration places velocities at the cell faces, and assigns pressure, stresses, and other material properties at the cell centers. We use a second-order central finite difference scheme for spatial discretization, except for the advection term arising from the upper convective derivative, where a fifth-order WENO (Weighted Essentially Non-Oscillatory) scheme is employed to enhance accuracy and stability~\cite{Shu2009}. The time advancement of all terms, except for the non-Newtonian extra stress tensor, is managed with a second-order Adams-Bashforth scheme coupled with a fractional step method~\cite{Kim1985}, while the extra stress tensor is advanced using the Crank-Nicolson scheme. A fast Poisson solver based on the Fast Fourier Transform (FFT) is used to ensure a divergence-free velocity field.  Parallelization of the solver is achieved through the use of the domain decomposition library \textit{2decomp} (\texttt{http://www.2decomp.org}) and the MPI protocol. The evolution equation for the extra EVP stress tensor is formulated using the log-conformation method~\cite{Izbassarov2018} to ensure the positive definiteness of the conformation tensor, thereby preventing numerical instabilities that commonly arise in simulations of complex non-Newtonian fluids.

\subsection{Flow setup} 

\begin{figure}[!h]
\includegraphics [width = 0.45\textwidth]{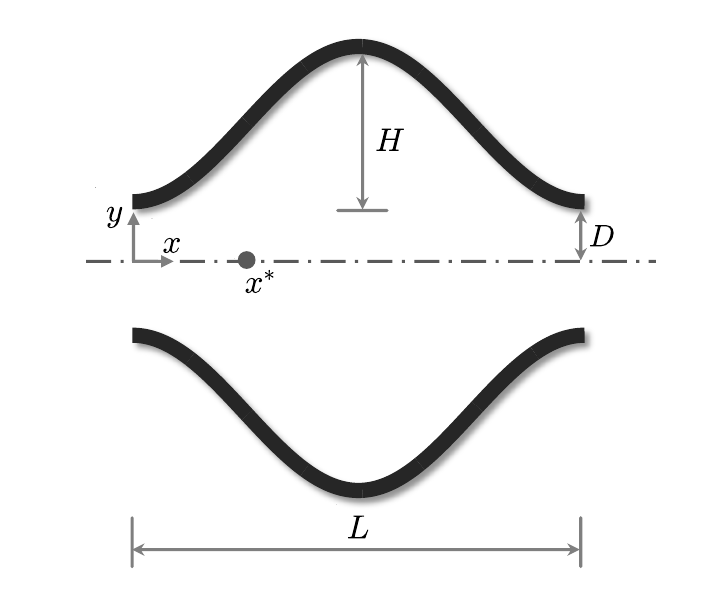}
\caption{\label{fig:wavy} Sketch of the flow domain of a single periodic wavy channel. The flow direction is from left to right. For specific analysis, we probe the flow statistics at a location $x^{*} = L/4$.}
\end{figure}

In this study, we investigate the flow of an incompressible EVP fluid through a single periodic wavy channel. The channel is illustrated in Fig. \ref{fig:wavy}. 
The numerical domain is a rectangle that measures $L \times 2(D+H)$. The top wall of the wavy channel follows $y(x) = D+0.5A (D - cos 2 \pi x /\lambda)$, while the bottom wall is symmetrically shaped. The aspect ratio, denoted by $A=H/D$, measures the relative depth of wall perturbations to the channel's half-width, while $\lambda=L/D$ quantifies the dimensionless wavelength. A periodic boundary condition is applied along the stream direction $x$, while a non-slip boundary condition is enforced on the channel walls.

We start the flow by setting the velocity field and the EVP stress tensor to zero at the beginning of the simulation. The flow is driven at a constant bulk velocity to maintain a constant Reynolds number, and the required pressure drop is calculated dynamically at each time step. 
We limit our study to a Reynolds number of $Re =0.01$, where inertial effects are minimal, and to a viscosity ratio of $\alpha = 0.88$. We focus on the effect of plasticity, represented by the Bingaham number ($Bn$), and the elasticity, represented by the Weissenberge number ($Wi$), thus systematically varying these parameters within the range of $0 \leq Wi \leq 10$ and $0 \leq Bn \leq 12$. 
Under conditions where both $Wi$ and $Bn$ are set to zero, the fluid exhibits a Newtonian-like behavior. Conversely, setting $Bn$ to zero while $Wi$ is greater than zero results in a pure viscoelastic fluid behavior. Similarly, with $Wi$ at a minimal value of $0.01$ and $Bn$ greater than zero, the behavior tends to be that of a viscoplastic fluid. When both parameters are nonzero, the resulting dynamics are characteristic of EVP fluids. Thus, these combinations of $Wi$ and $Bn$ allow us to understand the flow behavior under different conditions.
Our primary analysis focuses on a reference setup with an aspect ratio of $A=3$ and a dimensionless wavelength of $\lambda= 8$. Furthermore, to understand how channel geometry affects the flow dynamics, we also vary the aspect ratio ($A$) and the dimensionless wavelength ($\lambda$), with changes studied under specific flow conditions to assess their impact.

Preliminary simulations were conducted using different grid sizes of $D/16$, $D/32$, and $D/64$. Since no significant differences between the $D/32$ and $D/64$ configurations were revelead, a uniform constant grid size of $D/32$ in all directions is chosen. It is worth noting that, the same grid spacings was used in a previous study~\cite{DeVita2018}.

\subsection{\label{sec:code_valid}Experimental setup and code validation}

The numerical solver used in this study has been thoroughly tested and validated in various types of flows, such as single and multiphase flows, laminar and turbulent flows, and Newtonian and non-Newtonian fluids. In particular, the same or a similar solver has been applied in studies including particle suspensions and droplets in an EVP fluid~\cite{Izbassarov2018}, turbulent flows of an EVP fluid~\cite{Rosti2018, Izbassarov2021, Abdelgawad2023}, EVP flows in porous media~\cite{DeVita2018}, and sphere sedimentation in an EVP fluid~\cite{Sarabian2020,Sarabian2022}.

\begin{figure}
\includegraphics [width = 0.5\textwidth]{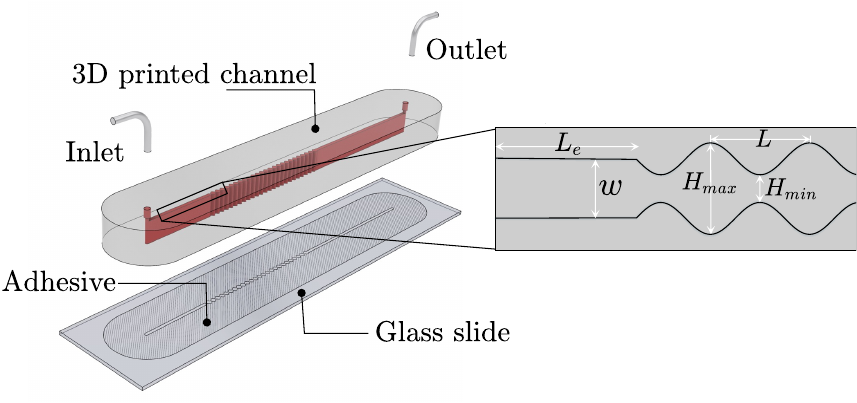}
\caption{\label{fig:channel_design} Schematic of the microfluidic channel used in the experiments. The channel has dimensions of $L_e =2$~cm, $w = 500~\mu$m, $H_{max} = 862~\mu$m, $H_{min} = 258~\mu$m, and $L = 800~\mu$m.}
\end{figure}

Additionally, we carry out micro-particle image velocimetry ($\mu$PIV) to analyze the flow dynamics of EVP fluids in a 2D wavy channel, allowing direct comparison between experimental velocity profiles and simulation predictions. The channel, crafted from fused silica glass via selective laser-induced etching~\cite{Burshtein2019}, measures 5~cm in length and 3~cm in depth, with an entry section of $L_e =2$~cm functions as a straight channel with a width of $w = 500~\mu$m, as depicted in Fig.~\ref{fig:channel_design}.
The wavy part of the channels spans 2~cm, including 25~periods, with widths varying from a minimum of $H_{min} = 258~\mu$m to a maximum of $H_{max} = 862~\mu$m and a wavelength of $L = 800~\mu$m. This configuration results in an aspect ratio of $A=2.3$ and a dimensionless wavelength of $\lambda=6.2$. The channel design incorporates an open side sealed against a transparent glass slide using UV adhesive, while the other side features two holes acting as inlet and outlet ports and glued to metallic fittings for tubing connections. The channel is connected to syringes through rigid polyethylene tubing to ensure minimal hydraulic compliance.

\begin{figure}
\centering
\includegraphics[width = 0.45\textwidth]{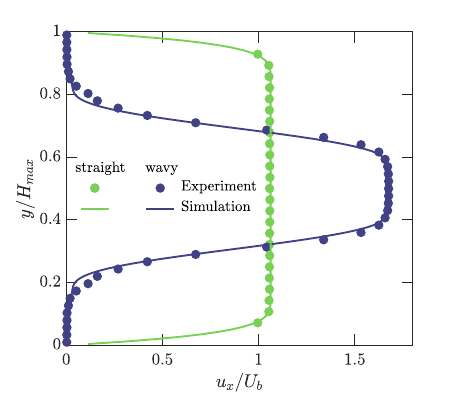}
\caption{\label{fig:valid_piv} Comparison of \textbf{experimental} (green) and \textbf{numerical} (blue) velocity profiles in straight and wavy channels at a flow rate of $Q = 0.5$~mL/min. $u_x$ is normalized by the bulk velocity $U_b$ in the horizontal axis, while the distance from the channel's bottom wall is normalized by its maximum width in the vertical axis.
}
\end{figure}

The EVP fluid used in the experiments is a Pluronic F127 (PF127) aqueous solution at a concentration of 21~wt.$\%$. Its rheological characteristics, including a yield stress of $\tau_y = 135$~Pa, a flow consistency index of $k = 54 $~Pa~s$^n$, a power index of $n = 0.28$, and an elastic modulus of $G' = 970$~Pa, are detailed in Ref.~\citenum{Abdelgawad2024}. This thermoresponsive gel behaves as a Newtonian fluid below its gelation temperature of $23.8^\circ$C ~\cite{Abdelgawad2024}.
In a cold environment at about $4^\circ$C, the PF127 solution is seeded with $2~\mu$m fluorescent polystyrene particles (PS-FluoRed-Particles; Microparticles GmbH) at a concentration of approximately 0.02~wt.$\%$ before loading into the channel through the attached tubing and syringes. The system is then moved to the experimental space set at $26^\circ$C, and kept there long enough to reach thermal equilibrium with the surroundings, ensuring the transition to a gel state before any measurements are taken.
The flow is driven at a constant flow rate by a syringe pump (neMESYS, Cetoni GmbH), driving two $5$~mL glass syringes (Hamilton, Gastight Glass) in a push-pull setup.
For flow visualization, an illumination system (TSI Inc.) is used, which is equipped with a 527~nm dual-pulsed Nd laser integrated into an inverted microscope (Nikon ECLIPSE Ti-S) with a 5X objective lens (Nikon PlanFluor). The channel is positioned so that the glass slide is perpendicular to the light source, with the channel mid-plane brought into focus. Flow images are recorded using a $1280 \times 800$~pixels high-speed CMOS camera (Phantom Miro M310, Vision Research Inc.), set to capture 50 pairs of images at a frame rate adjusted to the specific flow rate. The resulting images are processed using an ensemble average cross-correlation PIV algorithm (TSI Insight 4G) to obtain the velocity field.
Streamwise velocity profiles ($u_x$), measured through $\mu$PIV along the maximum width for both straight and wavy sections, are compared with simulation results obtained using the SRM/HB model. This comparison, illustrated in Fig.~\ref{fig:valid_piv}, shows a good agreement between the experimental data and the model predictions.
This validation is conducted at a flow rate of $Q = 0.5$ mL/min; however, further comparisons at varying flow rates and different sections of the wavy channel consistently exhibit good agreement (not shown), affirming the reliability of our numerical methods.
\begin{figure*}[!t]
\centering
\includegraphics[width = 0.9\textwidth]{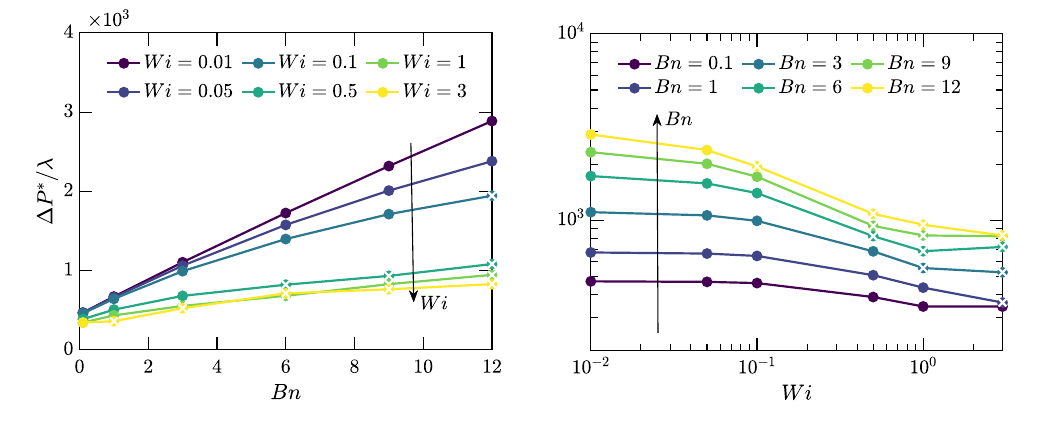}%
\caption{\label{fig:deltsP} Dimensionless mean pressure drop ($\Delta P^*/\lambda$) as a function of Bingham number ($Bn$, left), and  Weissenberg number ($Wi$, right), obtained from \textbf{numerical} simulations.}
\end{figure*}
\section{Results and discussion}
\subsection{Pressure drop}
We begin our analysis by focusing on how the Bingham and the Weissenberg numbers influence the pressure drop across the channel ($\Delta P/L$), a direct indicator of the drag experienced by the fluid. 
Fig.~\ref{fig:deltsP} illustrates the mean pressure gradient in its dimensionless form $\Delta P^*/\lambda$, where $\Delta P^ *= \Delta P/\rho U^2$, as a function of $Bn$ for various $Wi$ on the left, and as a function of $Wi$ for different $Bn$ values on the right. As $Bn$ increases, $\Delta P^*/\lambda$ similarly rises across all $Wi$, indicating a corresponding increase in drag. Notably, the effect of $Bn$ on $\Delta P^*/\lambda$ becomes less pronounced at higher $Wi$, suggesting that elasticity moderates the increase in drag caused by plasticity. Specifically, at very low $Wi$ values, such as $Wi = 0.01$, where elastic effects are negligible, $\Delta P^*/\lambda$ shows a steep and linear increase with $Bn$, typical of the viscoplastic behavior~\cite{Chaparian2020}. As soon as $Wi$ slightly increases to 0.05, this relationship deviates from linear.
For higher $Wi$ values (e.g., $Wi =1$ and $Wi =3$), the increase in $\Delta P^*/\lambda$ with $Bn$ is much more gradual, indicating a strong deviation from the viscoplastic case. These observations align with what has been reported by previous studies, i.e., that at finite elasticity, the plasticity of the fluid promotes the elastic behavior of the material, and the fluid behavior converges towards that of a viscoplastic fluid at lower $Wi$ when $Bn$ is large~\cite{Chaparian2019}. Notably, in some cases depending on the values of $Wi$ and $Bn$, the flow becomes time-dependent, exhibiting periodic or chaotic behaviors, as discussed later in Sec.~\ref{sec:time_depend}. We present the mean values for these time-dependent cases, marking them with an "$\times$" symbol in the corresponding figures.
Fig.~\ref{fig:deltsP} (left) shows that the effect of $Wi$ on $\Delta P^*/\lambda$ is more significant at higher $Bn$, demonstrating that fluids with higher yield stress benefit more from elastic properties in terms of drag reduction, see e.g. the significant decrease in $\Delta P^*/\lambda$ with $Wi$ at $Bn =12$. Conversely, for low $Bn$ (e.g., $Bn =0.1$), $\Delta P^*/\lambda$ remains almost constant, indicating a lesser impact of elasticity on the drag, which aligns with the behavior expected in viscoelastic flows in such geometries.

A similar effect of $Bn$ and $Wi$ on drag has been reported for laminar EVP flows in a straight channel~\cite{Izbassarov2021}. In this context, the friction factor increases with the $Bn$ and decreases with the $Wi$ number~\cite{Izbassarov2021}. This behavior is attributed to the size and distribution of the central solid plug region inside the channel, which moves at a uniform velocity and increases in size with $Bn$. A larger plug region volume results in a smaller yielded region, leading to a sharp shear rate near the walls and an increase in wall shear stress. In contrast, the size of the solid plug in the middle of the channel decreases with increasing $Wi$. Elasticity increases the total stress in the channel, causing more material to yield and reducing the wall shear stress, thus resulting in less drag~\cite{Izbassarov2021}. This suggests a correlation between the unyielded region volume and the pressure drop observed. We also perform straight channel simulations, detailed in the Appendix~\ref{app:straight_channel}, covering a broader range of $Wi$ and $Bn$ values, confirming these results and extending the findings reported in earlier studies. 

\subsection{Size and shape of unyielded regions}
\begin{figure*}[!t]
\centering
\includegraphics[width = 0.9\textwidth]{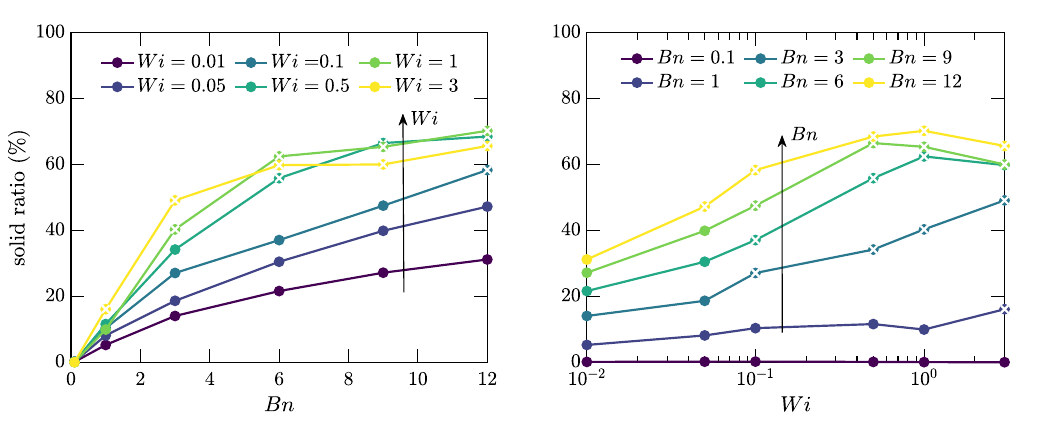}
\caption{\label{fig:SR} Ratio of unyielded regions volume to total flow domain volume as a function of Bingham number ($Bn$, left), and  Weissenberg number ($Wi$, right), obtained from \textbf{numerical} simulations.}
\end{figure*}

We calculate the ratio of the volume of the solid region to the entire flow domain of the wavy channel, shown in Fig.~\ref{fig:SR}
\begin{figure*}[!t]
\centering
\includegraphics[width = 0.92\textwidth]{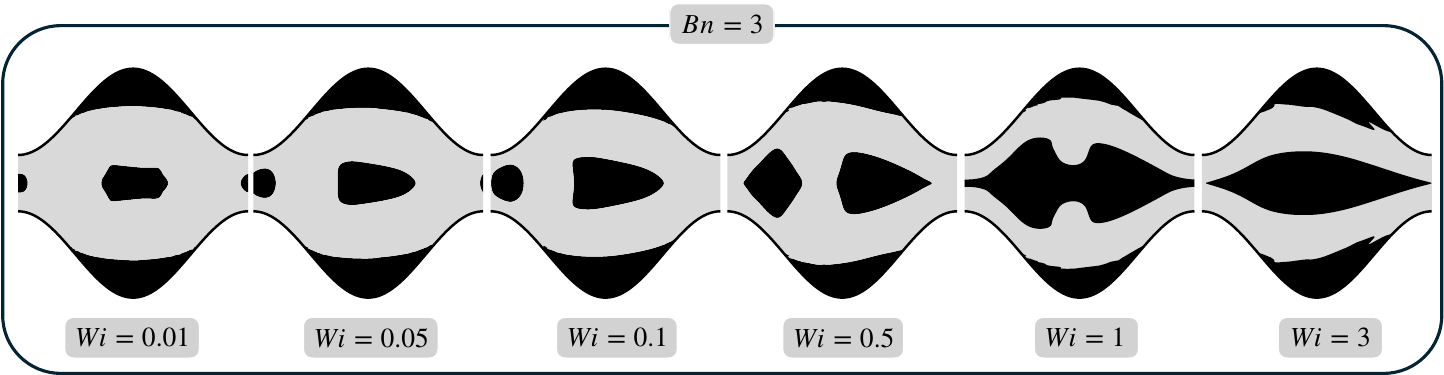}%
\caption{\label{fig:Yield_Unyield} Variations in yielded (gray) and unyielded (black) regions for changing $Wi$ at $Bn = 3$, obtained from \textbf{numerical} simulations.}
\end{figure*}
as a function of the Bingham ($Bn$) and Weissenberg ($Wi$) numbers. It is clear that the solid ratio increases with $Bn$ for all $Wi$ cases, while its dependence on $Wi$ increases with plasticity, similar to the observations in a straight channel. However, unlike in a straight channel, the elasticity here also increases the volume of the solid region, despite still reducing the drag.

To better understand the changes in the unyielded regions with elasticity, we examine the spatial variation of yielded and unyielded regions with $Wi$ at a given case of $Bn=3$, as depicted in Fig.~\ref{fig:Yield_Unyield}.
A key distinction between VP and EVP fluids lies in the ability of EVP fluids to undergo elastic deformations before yielding. This implies that in EVP, the rate of deformation tensor $\dot \gamma$ within the unyielded regions is not necessarily zero. To identify these unyielded regions, we set a threshold defined by the yield variable $Y = \max (0, 1-\tau_y/\tau_d) \le 0.01$, beyond which the size of the unyielded regions remains relatively constant.
The unyielded regions in Fig.~\ref{fig:Yield_Unyield} are shown in black, contrasting with the yielded regions in light gray.
In straight channels, the unyielded plug is typically located around the channel center, where the stress is minimal. However, introducing enough wall perturbations, such as waviness, creates stationary unyielded regions attached to the widest section of the channel~\cite{Roustaei2013}. The wall perturbations also induce elongational strain due to mass conservation, potentially yielding parts of the central plugs that can be continuous or fragmented depending on the flow geometry and conditions. While the unyielded regions along the walls remain stationary and serve as a lubrication layer, the central unyielded zones are not stagnant. Instead, they are merely low-stress regions where the yielded fluid particles decelerate and instantaneously solidify upon entering these zones, while liquefying and accelerating upon exiting~\cite{Varchanis2020}. Increasing the elasticity enlarges the unyielded regions, particularly the central ones, which become more streamlined and connected at high $Wi$, as shown in Fig.~\ref{fig:Yield_Unyield}. Notably, with increasing elasticity, these central regions exhibit pronounced right-left asymmetry, with respect to the vertical symmetric plane, even in the absence of inertial effects. This asymmetry has been previously reported in similar configurations, with some researchers suggesting that the asymmetry in the flow field and unyielded region shapes remains relatively constant for any given $Wi \times Bn$ combination~\cite{Chaparian2019, Villalba2023}.

In straight channels, the only nonzero components of the extra stress tensor are the shear stress ($\tau_{xy}$) and the first normal stress ($\tau_{xx}$). The flow is homogeneous along the streamwise direction, with no stress variations along the channel. This simplifies the presentation of stress profiles, as discussed in Appendix~\ref{app:straight_channel} and illustrated in Fig.~\ref{fig:straight_Bn9}. However, the scenario shifts dramatically in wavy channels, where wall undulations disturb the flow, leading to spatial variations in velocity and stresses, with a second normal stress ($\tau_{yy}$) being introduced. This complexity necessitates a detailed examination of stress distributions, which vary significantly along the channel due to introduced geometrical features. 
Therefore, to elucidate the impact of elasticity on the shape and size of the unyielded regions, we examine the stress fields within the wavy channel when varying $Wi$. 
\begin{figure*}[!t]
\centering
\includegraphics[width = 0.92\textwidth]{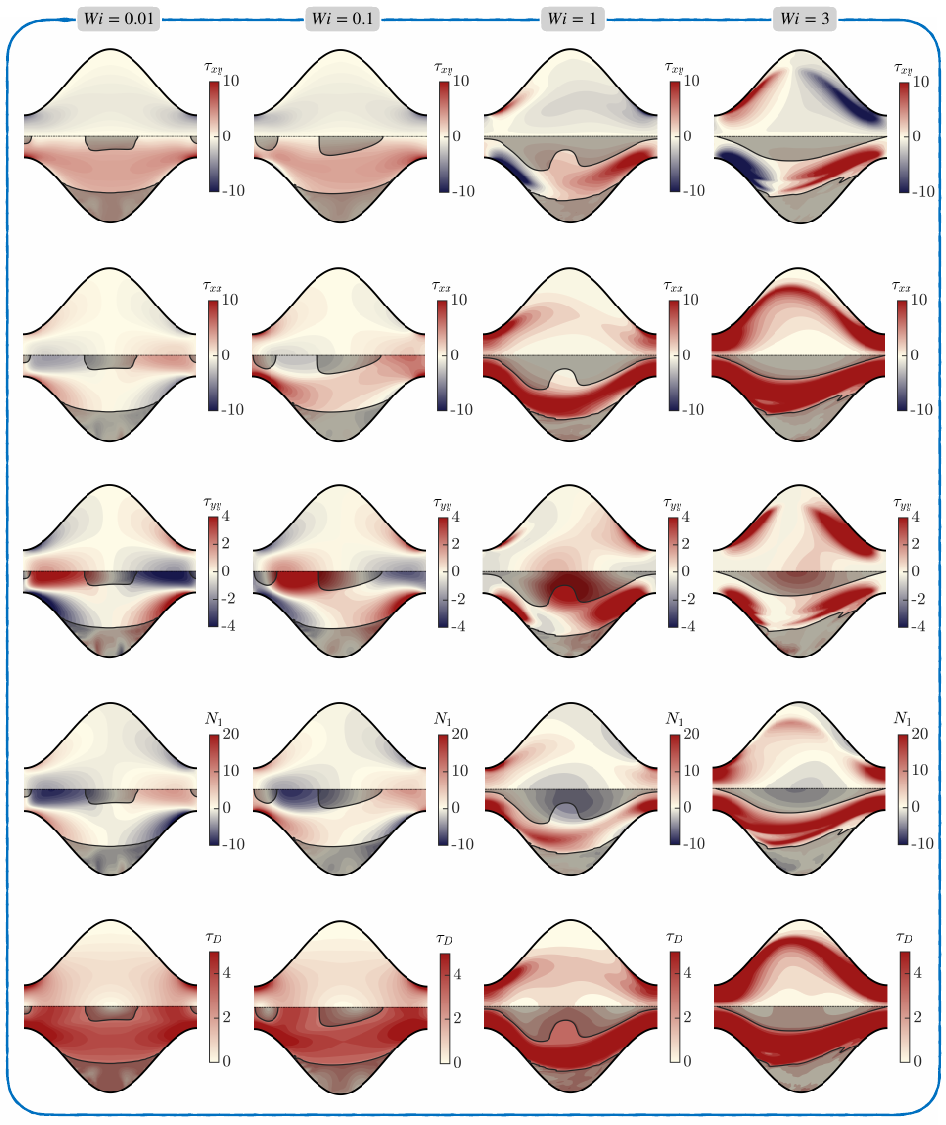}
\caption{\label{fig:VE_EVP} 
Comparison of the \textbf{numerical} time-averaged stress components between VE and EVP fluids at the same $Wi$, with the EVP cases at $Bn = 3$. $\tau_{xy}$, $\tau_{xx}$, $\tau_{yy}$, $N_1$, and $\tau_d$ are displayed from top to bottom, for VE fluids in the upper half and EVP fluids in the lower half of each panel. Unyielded regions in the EVP cases are highlighted with a transparent gray overlay.}
\end{figure*}

Fig.~\ref{fig:VE_EVP} reports a comparative representation of time-averaged stress components and their combination for the VE and EVP cases at different $Wi$, presented in columns, with the EVP cases at $Bn=3$. Each stress quantity is visualized for VE fluids in the top half of each panel, and for EVP fluids in the bottom half.
In the VE cases, stress fields at low elasticity ($Wi = 0.01$) appear almost symmetrical along the vertical axis. As elasticity increases, progressing from left to right, not only the magnitudes of all stress components increase, but the distribution of these stresses also becomes more asymmetric and localized, particularly close to the channel's narrowest sections. A similar trend is mirrored in the EVP scenarios, where the initial symmetry at low $Wi$ gives way to notable asymmetry as elasticity increases.
At low elasticity ($Wi = 0.01$), $\tau_{xx}$ and $\tau_{yy}$ both assume positive and negative values in a symmetric distribution, corresponding to the fluid navigating through the expansion and contraction parts of the channel. With an increase in $Wi$, e.g., to 0.1, a downstream shift in stress contours becomes evident, even at the low $Re$ considered here. 
This shift disrupts the symmetry within the stress components, leading to an asymmetric distribution of deviatoric stress $\tau_D$, shown in the bottom row of Fig.~\ref{fig:VE_EVP}. The asymmetry and localization of the stresses, coupled with the increasing elasticity, sharpen the stress gradients, prompting the yielding of central plugs around the channel's narrowest sections. Away from these localized high-stress zones, regions of low stress expand, allowing the central unyielded plugs to grow in size. This mechanism explains why unyielded regions not only expand, but also take a streamlined shape as elasticity increases.

\begin{figure*}[!t]
\centering
\includegraphics[width = 0.92\textwidth]{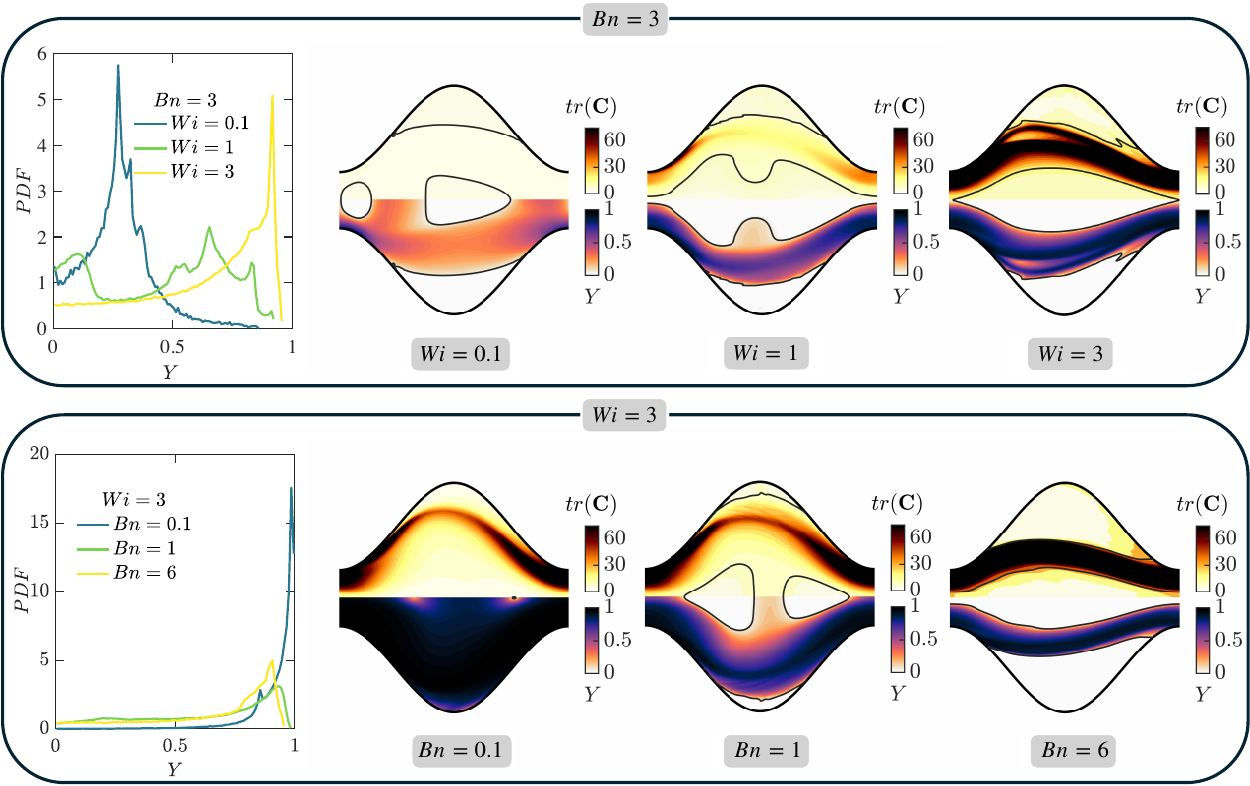}
\caption{\label{fig:trC_Y} Contour plots of the \textbf{numerical} time-averaged trace of the conformation tensor $tr(\boldsymbol C)$ in the upper half, and yield variable $Y$ in the lower half of each panel, for (top panel) different $Wi$ at $Bn = 3$ and for (bottom panel) different $Bn$ at $Wi = 3$. Each panel includes the probability distribution function (PDF) of $Y$ for each case.
}
\end{figure*}
\subsection{Yield variable and elasticity}
From Fig.~\ref{fig:VE_EVP}, it is clear that the yielded regions in EVP flows exhibit significantly higher stress values compared to their VE counterparts at the same $Wi$. To understand this, one should examine what the SRM model (Eq.~\ref{eq:evp_dimless}) predicts for the unyielded viscoelastic fluid and compare it to the classical Oldroyd-B model.
The main difference between the two constitutive models lies in the presence of the yield variable $Y \in [0,1]$, which represents the von Mises criterion. For the Oldroyd-B model, $Y$ is fixed equal to one, while in the SRM model, $Y$ varies locally between zero and one. In unyielded regions, $Y$ is zero, and in yielded regions, it ranges from zero to one. Dividing Eq.~\ref{eq:evp_dimless} by $Y$, an equivalent modified Oldroyd-B model for the viscoelastic yielded region is obtained,
\begin{equation} \label{eq:evp_dimless_Y}
	Wi^* ~\overset{\nabla}{\boldsymbol{\tau}^{*}}+\boldsymbol{\tau}^{*}=\alpha^* \boldsymbol{\gamma} ^{*},
\end{equation}
where $Wi^*$ and $\alpha^*$ are the modified Weissenberg number and viscosity ratio, respectively, scaled by the yield variable: $Wi^ *= Wi/Y$ and $\alpha^* = \alpha/Y$. In the yielded regions, where $Y \in (0,1]$ in the yielded region, the flow has a higher $Wi^* \ge Wi$ that changes locally depending on the value of the yield parameter at every point in the domain. 

To explore how the yield variable $Y$ affects the flow behavior in the yielded region, we examine its field and distribution at different $Wi$ and $Bn$ in Fig.~\ref{fig:trC_Y}. This analysis is linked to the trace of the conformation tensor $\boldsymbol{C} = \boldsymbol{\tau}/G + \boldsymbol{I}$, which is a measure of the stretching of the polymers. 
Increasing $Wi$ while maintaining a constant $Bn = 3$, $tr(\boldsymbol C)$ increases significantly within the yielded regions, correlating with the areas of high $Y$, as shown in Fig.~\ref{fig:trC_Y} (top panel).
As elasticity increases, the conformation tensor shows increased localization, mirroring the trends of the stress distributions, as shown in Fig.~\ref{fig:VE_EVP}. This suggests that the variation in the material stretching in unyielded regions intensifies with elasticity.

Analyzing the distribution of the yield variable $Y$ within the yielded region provides insights into its deviation from a classical Oldroyd-B viscoelastic model at corresponding $Wi$ values. This distribution is examined by calculating the probability distribution function (PDF), displayed in Fig.~\ref{fig:trC_Y} (top panel) left. At low $Wi$ ($Wi = 0.1$), $Y$ has a distribution limited to low values with a peak around $Y = 0.27$. As $Wi$ increases to 1 and further to 3, the distribution of $Y$ shifts towards higher and higher values, and becomes more peaked, showing a sharp maximum around $Y = 0.91$ for $Wi = 3$. This suggests that, at the same $Bn$, a flow with low elasticity is more likely to exhibit elastic effect compared to a viscoelastic fluid at the same $Wi$. 
The maximum elongation in the unyielded region also increases with elasticity, with  $tr(\boldsymbol C)_{max} \approx$ 3.1, 6.6, and 13.1 for $Wi=$ 0.1, 1, and 3, respectively. Indeed, the material ability to deform elastically before yielding increases with elasticity.
Increasing $Bn$ from 0.1 to 1, while keeping a constant $Wi = 3$, $tr(\boldsymbol C)$ exhibits minimal changes, yet significantly escalates when $Bn$ is increased to 6, as illustrated in Fig.~\ref{fig:trC_Y} (bottom panel). 
The distribution of the yield variable varies significantly across different Bingham numbers. At $Bn = 0.1$, the PDF peaks sharply near $Y= 0.98$, indicating a similarity to an Oldroyd-B viscoelastic fluid at low plasticity. As $Bn$ increases to 1 and then to 6, the peak of the PDF shifts towards lower values, and the distribution broadens. This suggests that with higher plasticity, the behavior of the yielded fluid increasingly deviates from that of a typical Oldroyd-B viscoelastic fluid. In other words, at high plasticity, the yielded regions of the EVP fluid has, on average, a much higher ratio of $Wi^*/Wi$ than at low plasticity. For instance, $Wi^*/Wi \approx 12.6$ at $Bn=0.1$, while  $Wi^*/Wi \approx 4.4$ at $Bn=6$. Overall, the presence of the yield variable in the SRM model enhances the elasticity of yielded regions, effectively coupling elasticity and plasticity characteristics in this framework.

\subsection{\label{sec:time_depend}Time-dependent behavior}
\begin{figure}
\centering
\includegraphics[width = 0.45\textwidth]{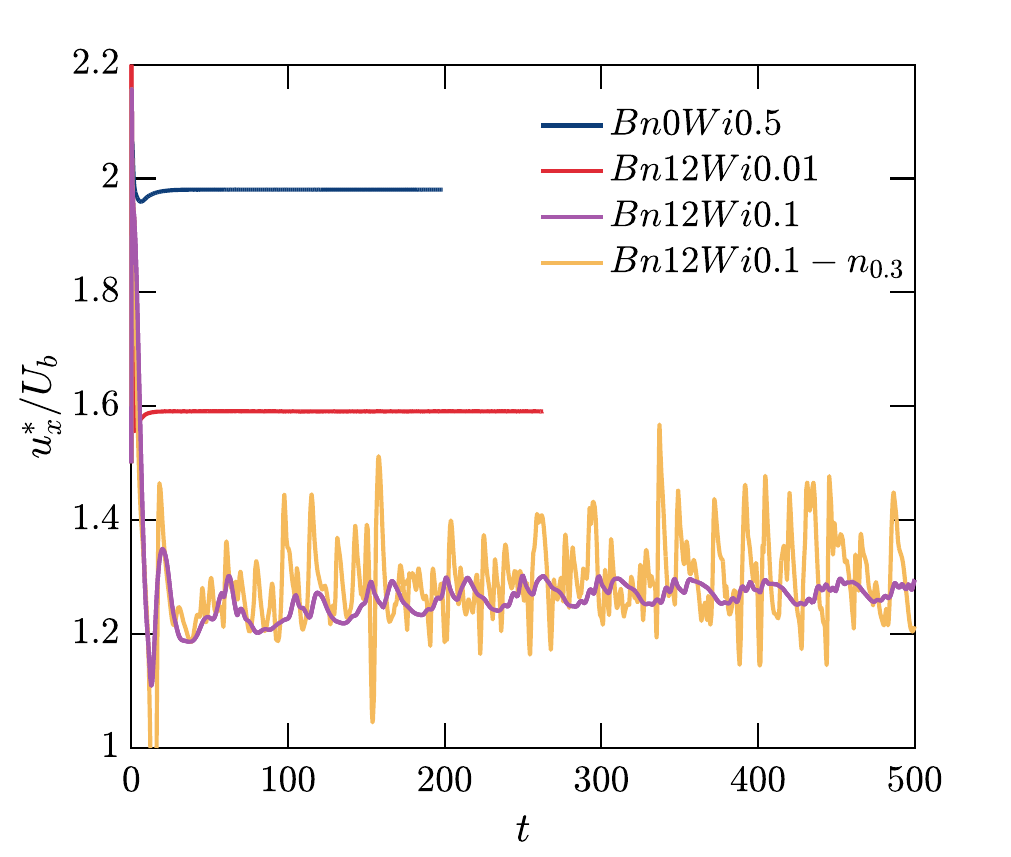}
\caption{\label{fig:timeDepend} Time history of the \textbf{numerical} streamwise velocity component $u_{x^*}$, for different cases, evaluated at $x^*$, see Fig.\ref{fig:wavy}.}
\end{figure}

Recent studies have shown that EVP fluids exhibit a time-dependent behavior at high $Wi$~\cite{DeVita2018,Parvar2024}, and, interestingly, this behavior can manifest at much lower levels of elasticity compared to pure viscoelastic fluids~\cite{CHAUHAN2024,MOUSAVI2024}, suggesting a special sensitivity of EVP fluid to elasticity. 
Here, we observe a similar phenomenon in EVP flow through a wavy channel. For example, with $Bn = 12$, the onset of transient behavior occurs at $Wi = 0.1$, while the VE flow remains steady at $Wi = 0.5$, as illustrated in Fig.~\ref{fig:timeDepend}.
The figure shows the time evolution of the streamwise velocity component probed at position $x^* = L/4$ and $y = 0$, which corresponds to the midpoint of the entrance to the contraction part. 
While viscoplastic flows do not exhibit any transient behavior, as is evident in the case of $Bn=12$ and $Wi=0.01$, the threshold in elasticity necessary for triggering an unsteady behavior depends significantly on the fluid plasticity. For example, at $Bn =12$, fluctuations in the streamwise velocity component become noticeable at $Wi=0.1$, while at a lower $Bn$, e.g. $Bn = 3$, a $Wi$ larger than 1 is required. This observation suggests that the onset of time-dependent flow behaviors is not simply governed by the product $Bn \times Wi$, as suggested in the past~\cite{CHAUHAN2024}. Instead, it involves a complex interplay between elasticity and plasticity, and flow geometry as will be shown later. Moreover, the presence of shear thinning can accentuate the time-dependent behaviors, as shown for the case of $Bn=12$ and $Wi=0.01$ with a power index of $n=0.3$.

\begin{figure}
\centering
\includegraphics[width = 0.45\textwidth]{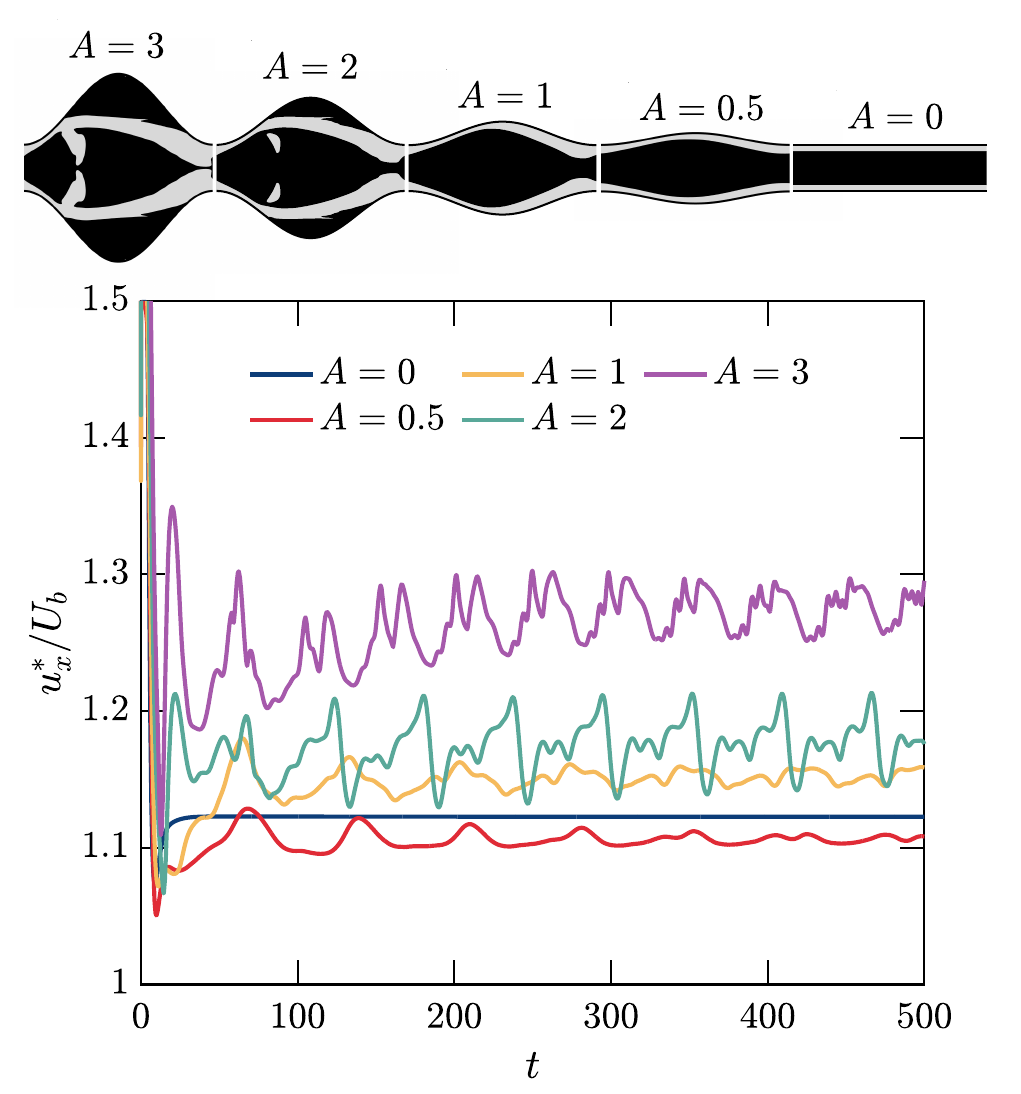}
\caption{\label{fig:shape_AR_time_depend} Time history of the \textbf{numerical} streamwise velocity component $u_{x^*}$ for the case of $Bn=12$ and $Wi=0.1$ and different aspect ratios A of the wall perturbation. The time-averaged distribution of (gray) yielded and (black) unyielded regions is shown on top.}
\end{figure}

\begin{figure}
\centering
\includegraphics[width = 0.45\textwidth]{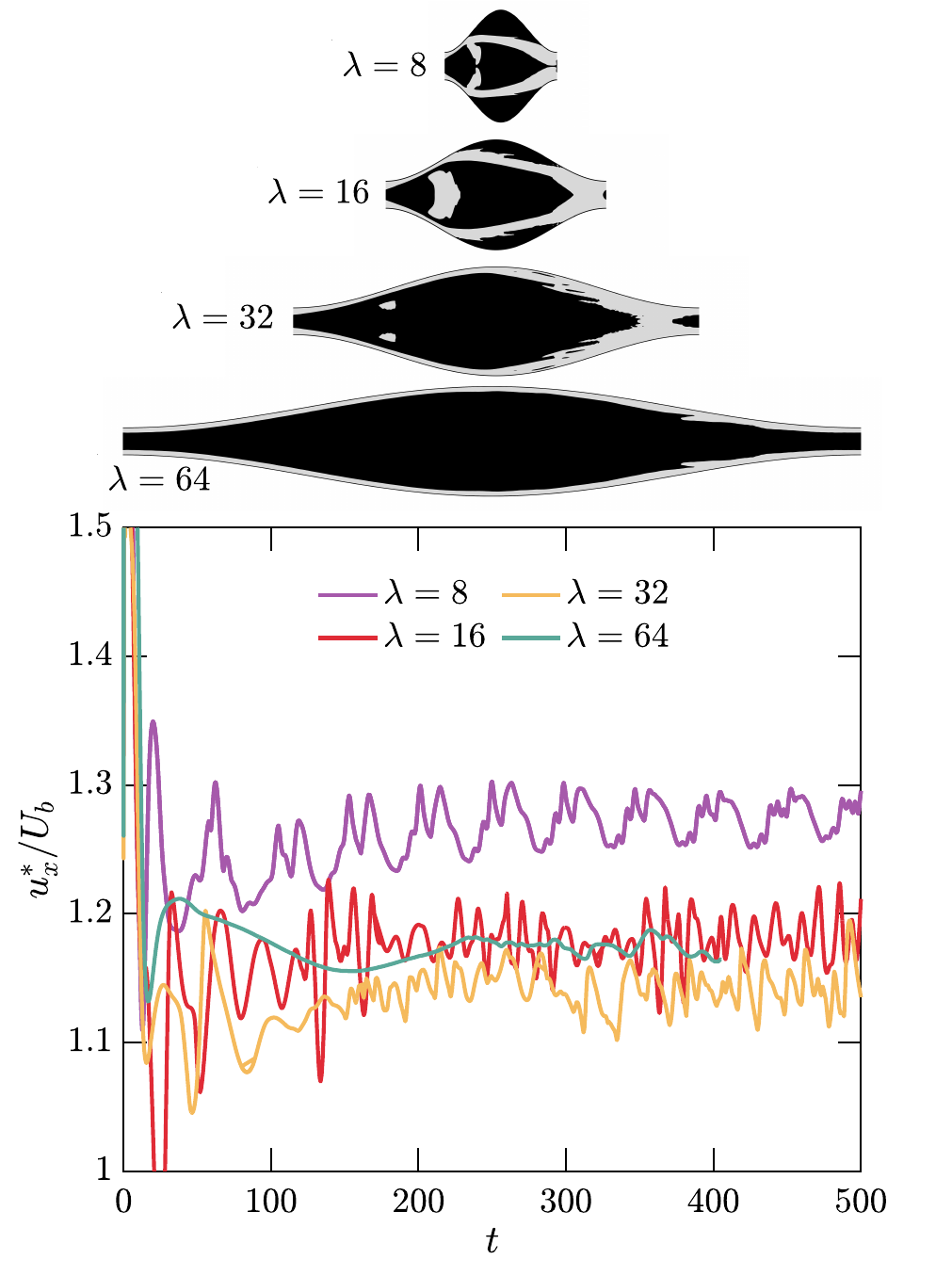}
\caption{\label{fig:shape_lambda_time_depend} Time history of the \textbf{numerical} streamwise velocity component $u_{x^*}$ over time for the case of $Bn=12$ and $Wi=0.1$ and different dimensionless wavelength $\lambda$. The time-averaged distribution of (gray) yielded and (black) unyielded regions is shown on top.}
\end{figure}

We further investigate the effect of the geometry on the flow regime by examining different aspect ratios and wavelengths of the wall ondulation. Focusing on the case with $Bn=12$ and $Wi=0.1$, which marks the onset of the time-dependent behavior in the reference geometry, we find that reducing the aspect ratio $A$, leads to a damping of the time fluctuations, with the flow approaching the steady state when $A$ equals zero, as depicted in Fig.~\ref{fig:shape_AR_time_depend}. Changing the aspect ratio also affects both the shape and distribution of the unyielded region within the wavy channel: as the aspect ratio decreases from 3 to 2, there is a noticeable reduction in the size of the stationary fouling layer, with the central plug zone appearing more connected; further reduction of $A$ causes the fouling layer to vanish, leaving only a central plug region that becomes increasingly symmetric with respect to the vertical central axis. At $A=0$, representing a straight channel configuration, the flow forms a uniformly straight central plug.
The changes of the dimensionless wavelength $\lambda$ of the channel similarly impact the transient behavior of the flow, as shown in Fig.~\ref{fig:shape_lambda_time_depend}. At large $\lambda$, e.g. $\lambda= 64$, the flow exhibits low-frequency fluctuations, which are expected to disappear as $\lambda \to \infty$ (a straight channel). Additionally, variations in $\lambda$ modify the shape and distribution of the unyielded regions across the channel in a similar fashion observed with varying aspect ratios, moving towards a straight channel design.
%

\subsection{Flow of highly elastic EVP}
\begin{figure}
\centering
\includegraphics[width = 0.45\textwidth]{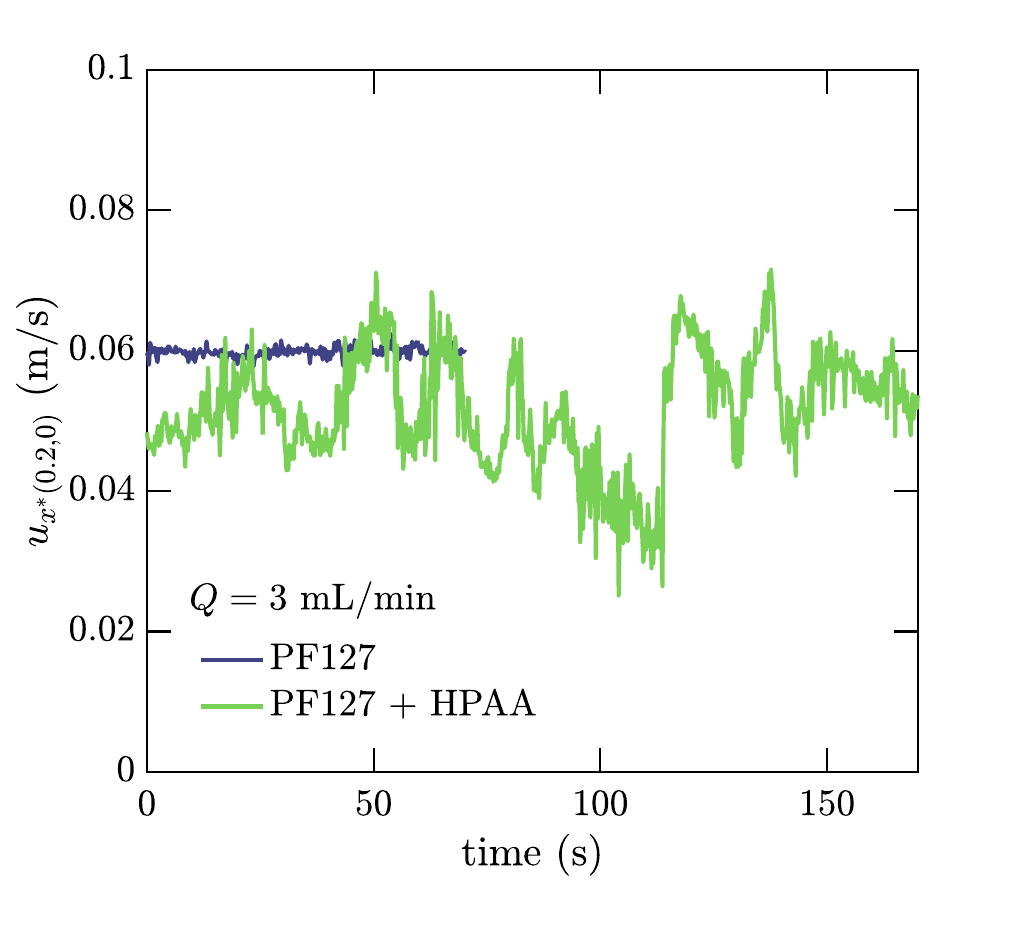}
\caption{\label{fig:ux(t)_3mlm} Comparison between the \textbf{experimental} streamwise velocity component $u_{x^*}$ measured by $\mu$PIV for PF127 and PF127+HPAA, at a flow rate of $Q = 3$~mL/min.}
\end{figure}

\begin{figure}
\centering
\includegraphics[width = 0.45\textwidth]{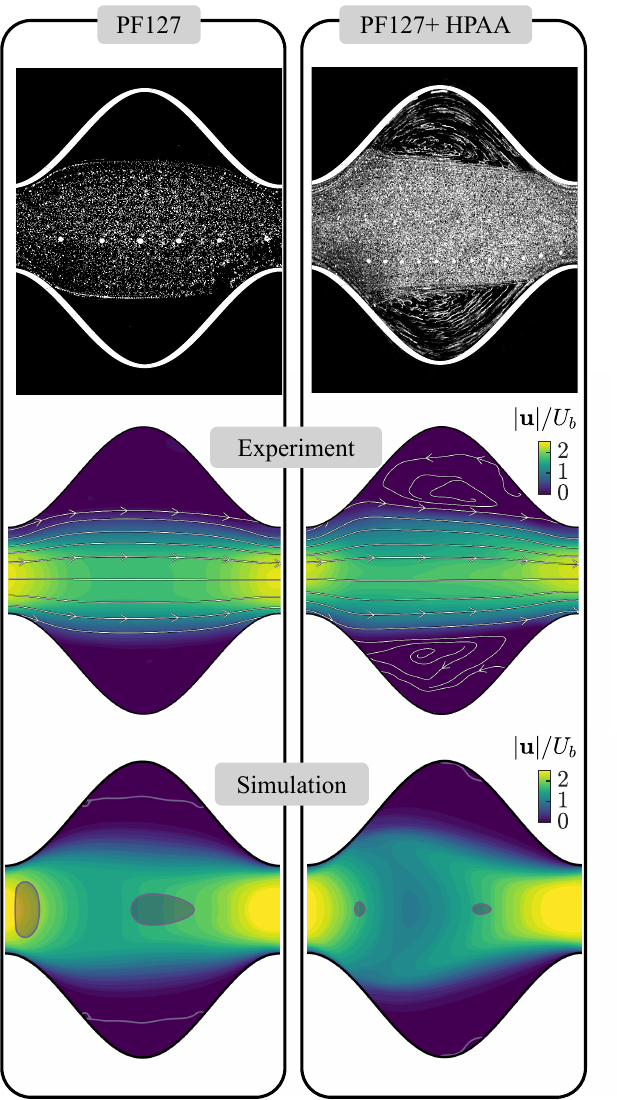}
\caption{\label{fig:exp_sim_3mlm} Comparison between the flow fields of (left column) PF127 and (right column) PF127+HPAA, at a flow rate of $Q = 3$~mL/min. Top raw: tracer particles pathlines. Middle raw: $\mu$PIV flow field and streamlines. Bottom raw: simulated flow field, with the unyielded region highlighted.}
\end{figure}

\begin{figure}
\centering
\includegraphics[width = 0.45\textwidth]{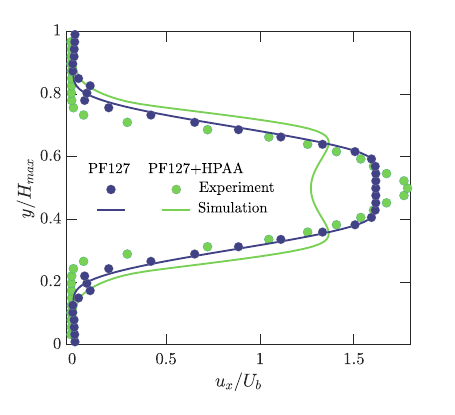}
\caption{\label{fig:ux_exp_sim_3} Comparison of \textbf{experimental} and \textbf{numerical} velocity profiles of PF127 and PF127+HPAA, at a flow rate of $Q = 3$~mL/min.}
\end{figure}

Despite the predicted time-dependent behavior of EVP fluids at high $Wi$, such dynamics have not been experimentally observed in low Reynolds number flows yet. This discrepancy stems from the fact that model EVP fluids used in flow experiments, such as Carbopol solutions, are weakly elastic, despite it being finite~\cite{Gouamba2019, Fraggedakis2016}. 
In a recent experimental study, we established that the introduction of polymeric additives, like hydrolyzed polyacrylamide (HPAA), to a model EVP fluid (PF127 21 wt.$\%$) can significantly enhance its extensional properties, such as relaxation times and extensional stresses, without altering its shear rheology~\cite{Abdelgawad2024}. This allowed us to design a model EVP fluid with a tunable elasticity, mimicking those in industrial settings. This modification promotes the fluid to exhibit elastic instabilities and time-dependent behavior under planar elongational flows, not observed before for EVP fluids~\cite{Abdelgawad2024}.
Here, we conduct experiments using the modified EVP fluid to examine its flow regimes in our wavy channel setup, for different $Wi$. In particular, we compare the flow behavior of pure PF127 21~wt.$\%$ (referred to as PF127) against the PF127 enhanced with 0.05 wt.$\%$ HPAA (referred to as PF127+HPAA): both fluids maintain the same shear rheology detailed in Sec.~\ref{sec:code_valid}, but PF127+HPAA is extensible, with a measurable relaxation time of 2.68~seconds~\cite{Abdelgawad2024}. The streamwise velocity, $u_{x}$, measured at $x^* = L/4 = 0.2$~mm and $y = 0$, is recorded over time for both fluids at a constant flow rate of $Q = 3$ mL/min. Our $\mu$PIV measurements reveal that PF127+HPAA indeed exhibits a pronounced time-dependent behavior, in contrast to pure PF127, which remains fairly steady within the level of our experimental measurement noise, as shown in Fig~\ref{fig:ux(t)_3mlm}.
This experimental observation provides first evidence of transient behaviors, potentially occurring in real-world scenarios when EVP fluids flow at high $Wi$. Interestingly, the flow fields for both cases are quite distinct, as shown in Fig~\ref{fig:exp_sim_3mlm}.
The flow of PF127 is typical, characterized by stationary zones formed at the widest section of the channel and a central plug zone surrounded by sheared regions separating the two zones. However, the flow of PF127+HPAA is characterized by having no stationary regions, showing instead a circulation regions near the wall, thus suggesting that the flow is no longer unyielded there. The stationary zones of PF127 and the circulation regions of PF127 with 0.05~wt.$\%$ HPAA flows are clearly visualized by the tracer particles pathlines reported in Fig.~\ref{fig:exp_sim_3mlm}.
These images are generated by time-averaging the pixel intensity over a number of consecutive frames for each case after removing the background for better clarity. In this way, stationary regions, where particles remain stagnant, appear as empty black zones, while regions where particles are circulating are highlighted with visible pathlines. Notably, the central plug zone almost does not exist in the PF127+HPAA case, as evidenced by the velocity profile along the central vertical axis shown in Fig.~\ref{fig:ux_exp_sim_3}.
The absence of stationary zones in the case of PF127+HPAA is attributed to the significant increase of normal stresses, which are generated at high $Wi$. This leads to global yielding and a flow behavior similar to that of viscoelastic fluids, as also observed in planar elongational flows~\cite{Abdelgawad2024}. We run simulations for both fluid cases, considering $Wi$ based on the relaxation time obtained from the extensional rheometry for PF127+HPAA. 
While the simulation results for PF127 closely match the experimental ones, the simulated flow field for PF127+HPAA, characterized by a smaller unyielded region, provides only a qualitative match but does not fully align with the experimental data, particularly when comparing flow fields and velocity profiles in Fig.~\ref{fig:ux_exp_sim_3}. This discrepancy is probably due to the fact that the SRM/HB model does not discern between shear and extensional material properties. Given that the PF127+HPAA fluid exhibits extensional characteristics not directly linked to its shear rheology, deviations between the modeled and actual flow behaviors can be expected when using Saramito's framework.

\begin{figure}
\centering
\includegraphics[width = 0.45\textwidth]{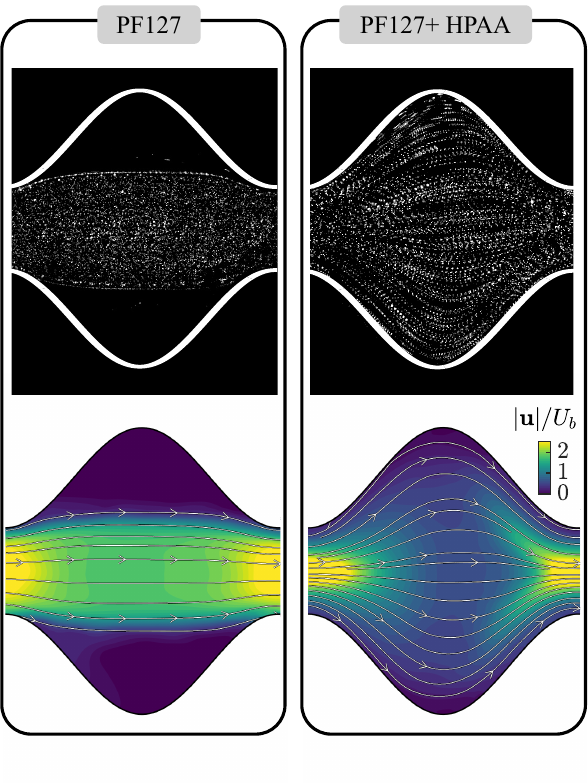}
\caption{\label{fig:exp_sim_1mlm} Comparison between the \textbf{experimental} flow fields of (left column) PF127 and(right column) PF127+HPAA, at a flow rate of $Q = 1$~mL/min. Top raw: tracer particles pathlines. Middle raw: $\mu$PIV flow field and streamlines.}
\end{figure}

Examining the flow of PF127 and PF127+HPAA at different flow rates reveals consistent qualitative behavior at low flow rates, characterized by stationary zones and central plugs. However, at sufficiently high flow rates, i.e., $Q \ge 0.5$ mL/min, PF127+HPAA exhibits a viscoelastic-like behavior with no stagnation regions at all. For example, at a flow rate of $Q = 1$~mL/min, the EVP flow of PF127+HPAA is completely yielded, fully invading the channel, displacing the existing fluid in the channel and replacing it with a new batch, as shown in Fig.~\ref{fig:exp_sim_1mlm}. 
This global yielding behavior observed experimentally with PF127+HPAA has significant implications for applications that involve the displacement of yield stress fluids, such as mud circulation in oil well cementing processes. 
Indeed, the potential ability of highly elastic EVP fluids to navigate complex geometries, such as wavy channels and porous media, presents numerous potential applications in areas requiring enhanced mixing and cleansing. 

\section{Conclusion}

In this study, we investigate the complex behavior of elastoviscoplastic (EVP) fluids in wavy channels using numerical simulations complemented by microfluidic experiments. We used the Saramito constitutive equation to model the EVP fluid and to investigate the interplay between elasticity and plasticity in such flows. Our study revealed that even minimal levels of elasticity can significantly alter the flow regime, pressure drop, and variations in yielded and unyielded regions within the channel.

We observe that elasticity can modulate the increase in pressure drop induced by high plasticity, with the effect of elasticity on reducing the pressure drop becoming more pronounced at higher levels of plasticity. Both the Bingham number, representing plasticity, and the Weissenberg number, representing elasticity, significantly impact the size and shape of the unyielded regions. An increase in the Bingham number typically enlarges these regions, while high Weissenberg numbers not only expand them but also introduce pronounced right-left asymmetry, despite negligible inertial effects. At high Weissenberg numbers, elasticity intensifies and localizes stresses along the channel's narrowest sections and streamlines, resulting in an enlarged central plug of a more streamlined shape. The presence of the yield variable, represented by the \textit{max} term in the Saramito model, significantly enhances the elasticity of yielded regions. The enhanced elasticity, which increases with plasticity, coupled with the stress distribution across the channel, allows EVP fluids to exhibit time-dependent behaviors at much lower Weissenberg numbers compared to viscoelastic fluids.

An important aspect of our study involved the examination of the flow behavior of a modified EVP fluid with enhanced elasticity; Pluronic F127 (PF127) with added hydrolyzed polyacrylamide (HPAA) polymeric additive. Our results show that this fluid displays unique behaviors not observed in less elastic fluids, such as pure PF127, even though both fluids share the same shear rheology. At high flow rates, this modified fluid exhibits global yielding and behaves similarly to viscoelastic fluids due to significant increases in normal stresses. The enhanced elasticity in PF127+HPAA enables the manifestation of time-dependent dynamics, behavior absent in the less elastic pure PF127 counterpart.
The ability of this highly elastic EVP fluid to exhibit such dynamic and yielding behaviors in complex geometries, such as wavy channels, underlines its potential for applications involving the displacement of yield stress fluids, such as in oil well cementing processes where effective mud circulation and displacement are crucial. The transient nature of the flow, facilitated by enhanced elasticity, could lead to improved mixing and cleansing effects in various industrial processes.

\begin{acknowledgments}
The research was supported by the Okinawa Institute of Science and Technology Graduate University (OIST) with subsidy funding to M.E.R. and A.Q.S. from the Cabinet Office, Government of Japan. The authors acknowledge the computer time provided by the Scientific Computing and Data Analysis section of the Core Facilities at OIST, and the computational resources on SQUID provided by the Cybermedia Center at Osaka University through the HPCI System Research Project (project ID: hp230018). S.J.H. acknowledges funding from the Japan Society for the Promotion of Science (JSPS) Kakenhi grants no. 24K07332 (S.J.H)), and 24K00810 (A.Q.S, M.E.R, S.J.H).
\end{acknowledgments}

\section*{Data Availability Statement}

The data that support the findings of this study are available within the article.

\appendix

\section{\label{app:straight_channel}Straight channel}
In this appendix, we present our results in a straight channel, focusing on the effect of $Bn$ and $Wi$ on the pressure drop and the volume ratio of the unyielded regions. Similar to our wavy channel simulations, the flow is driven at a constant flow rate. 

\begin{figure*}[!t]
\centering
\includegraphics[width = 0.9\textwidth]{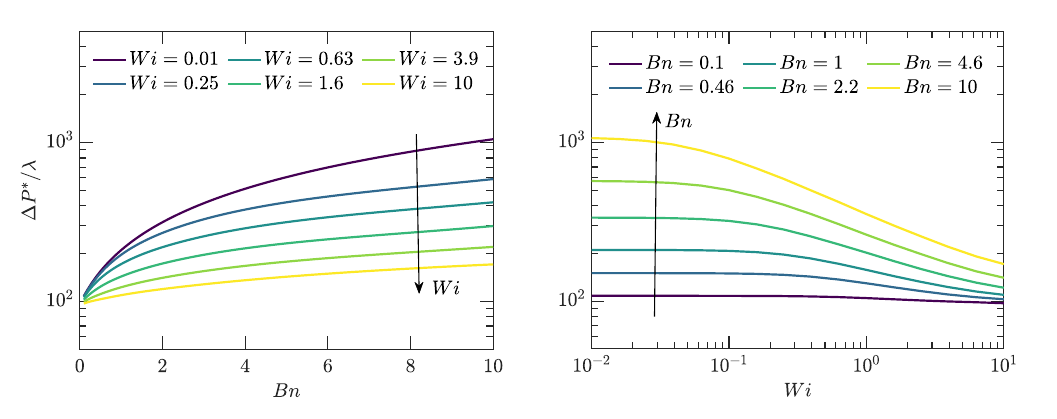}
\caption{\label{fig:straight_dP} Dimensionless pressure drop $\Delta P^*/\lambda$ as a function of (left) $Bn$ and (right) $Wi$ in a straight channel, obtained from \textbf{numerical} simulations.}
\end{figure*}

\begin{figure*}[!t]
\centering
\includegraphics[width = 0.9\textwidth]{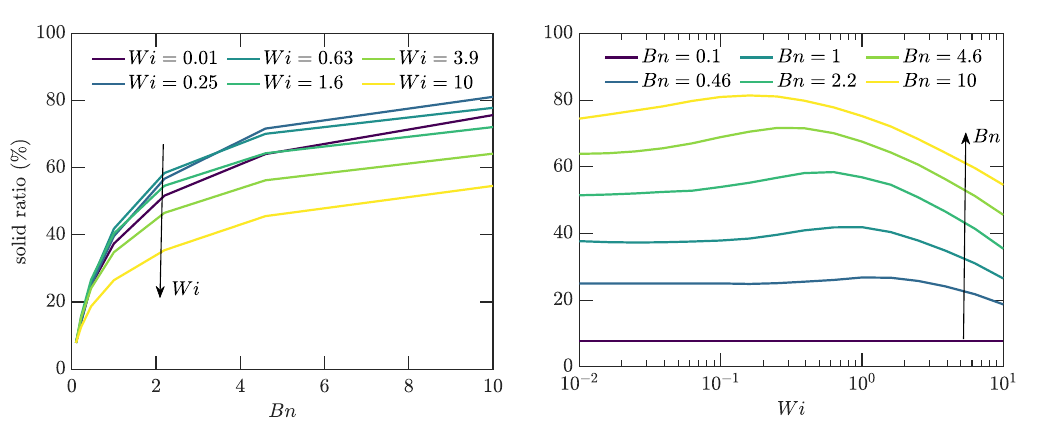}
\caption{\label{fig:straight_SR} Ratio of unyielded regions volume to the total flow domain volume, as a function of (left) $Bn$ and (right) $Wi$ in a straight channel, obtained from \textbf{numerical} simulations.}
\end{figure*}

\begin{figure*}[!t]
\centering
\includegraphics[width = 0.9\textwidth]{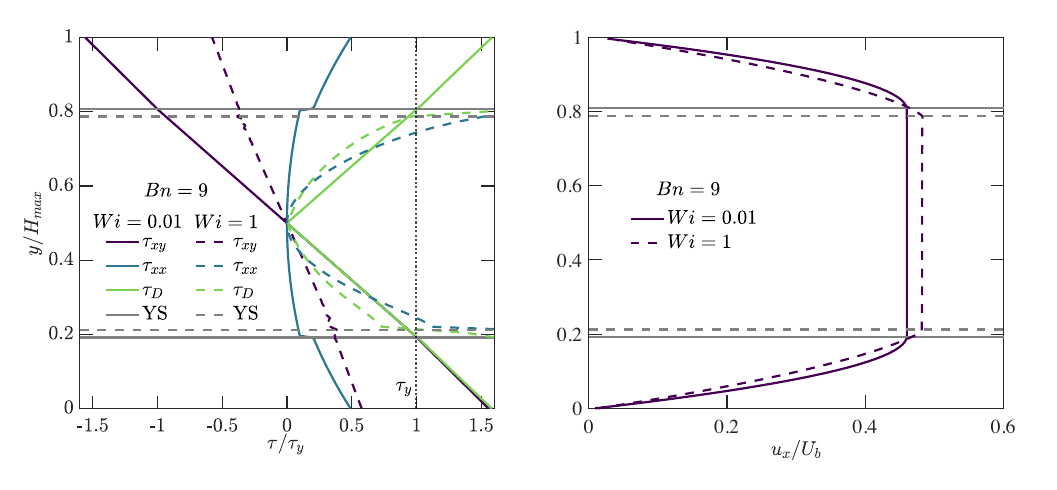}
\caption{\label{fig:straight_Bn9} Comparison of the profiles of (left) the extra stress components and (right) velocity, between two cases of the same $Bn=3$ and different $Wi$ equal to 0.01, and 1, obtained from \textbf{numerical} simulations.}
\end{figure*}

Fig.~\ref{fig:straight_dP} shows the dimensionless pressure drop as a function of $Wi$ and $Bn$. At low $Wi$, in the limit of viscoplastic fluid, the pressure drop remains constant before it starts to decrease when $Wi$ increases further. The onset of this reduction in pressure drop varies with $Bn$, occurring at lower $Wi$ values as $Bn$ increases. This indicates that the behavior of EVP fluids converges towards that of VP fluids at lower $Wi$ for higher $Bn$. At low $Wi$, pressure drop increases linearly with $Bn$, while its dependence on $Wi$ decreases as $Wi$ increases. The volume of the unyielded region increases with $Bn$, as the fluid becomes more plastic, as shown in Fig.~\ref{fig:straight_SR}. This volume initially also shows a plateau at low $Wi$ followed by a local maxima before it decreases as $Wi$ increases further. The value of $Wi$ marking the onset of this decrease depends on $Bn$, with its value increasing as $Bn$ grows, similar to the trend observed for the pressure drop. However, the onset of the decrease of the volume of the unyielded region is shifted towards higher $Wi$ values compared to that of the pressure. 
Beyond the onset value, both pressure drop and the volume of the unyielded region decrease with $Wi$, in a correlated manner. To better understand this correlation, we compare the profiles of the extra stress tensor components and the velocity for two cases at the same $Bn$ ($=9$), but at different $Wi$ ( 0.01 and 1), reported in Fig.~\ref{fig:straight_Bn9}. In straight channels, the only nonzero components of the extra stress tensor are the shear stress $\tau_{xy}$ and the first normal stress $\tau_{xx}$. The flow is homogeneous along the streamwise direction, with no stress variations, simplifying the presentation of the stress profiles. At low elasticity ($Wi = 0.01$), the shear stress is the dominant component and has a much higher absolute value than the normal stress, making it the main contributor to the magnitude of the deviatoric stress tensor $\tau_{D}$. The yield surface (YS) is defined through the intersection between $\tau_{D}$ and $\tau_{y}$, shown as a vertical line in the left panel of Fig.~\ref{fig:straight_Bn9}. At a higher elasticity ($Wi = 1$), $\tau_{xy}$ decreases while the normal stress $\tau_{xx}$ significantly increases, contributing to the increase in $\tau_{D}$. This increase in $\tau_{D}$ due to elasticity results in a decrease in the size of the unyielded region defined by the yield surface. The right panel of Fig.~\ref{fig:straight_Bn9} shows that the decrease in size of the unyielded region affects the velocity gradient in the sheared region between the wall and the central plug, leading to a smaller shear rate and, consequently, smaller wall shear stress and drag.

\appendix 

\nocite{*}

\bibliographystyle{unsrt}

%


\end{document}